# The Progenitor and Central Engine of a Peculiar GRB 230307A

ZhaoWei Du[1], HouJun Lü[1], Yong Yuan[2], Xing Yang[1], and EnWei Liang[1]
[1] Guangxi Key Laboratory for Relativistic Astrophysics, Department of Physics, Guangxi University, Nanning 530004, People's Republic of China; lhj@gxu.edu.cn
[2] School of Physics Science and Technology, Wuhan University, No. 299 Bayi Road 0, Wuhan, 430072 Hubei, People's Republic of China

## Abstract

Recently, a lack of supernova-associated with long-duration gamma-ray burst (GRB 230307A) at such a low redshift $z = 0.065$, but associated with a possible kilonova emission, has attracted great attention. Its heavy element nucleosynthesis and the characteristic of soft X-ray emission suggest that the central engine of GRB 230307A is a magnetar that is originated from a binary compact star merger. The calculated lower value of $\varepsilon \sim 0.05$ suggests that GRB 230307A seems to have an ambiguous progenitor. The lower value of $f_{\rm eff} = 1.23$ implies that GRB 230307A is not likely to be from the effect of "tip of iceberg." We adopt the magnetar central engine model to fit the observed soft X-ray emission with varying efficiency and find that the parameter constraints of the magnetar falls into a reasonable range, i.e., $B < 9.4 \times 10^{15}$ G and $P < 2.5$ ms for $\Gamma_{\rm sat} = 10^3$, and $B < 3.6 \times 10^{15}$ G and $P < 1.05$ ms for $\Gamma_{\rm sat} = 10^4$. Whether the progenitor of GBR 230307A is from the mergers of neutron star–white dwarf (NS–WD) or neutron star–neutron star (NS–NS) remains unknown. The difference of GW radiation between NS–NS merger and NS–WD merger may be a probe to distinguish the progenitor of GRB 230307A-like events in the future.

*Unified Astronomy Thesaurus concepts:* Gamma-ray bursts (629)

## 1. Introduction

Despite decades of studying gamma-ray bursts (GRBs) with much progress, several open questions regarding the progenitors, central engines, and their possibly associated gravitational waves (GWs) still exist in GRB physics (Zhang & Mészáros 2001; Kumar & Zhang 2015).

Phenomenally, GRBs are divided into two categories, which are long duration with soft spectrum and short duration with hard spectrum, with a division line at the observed duration $T_{90}$ 2 s (Kouveliotou et al. 1993). Observations of intense star formation in irregular galaxies and robust evidence associated with supernovae of some long-duration GRBs (Woosley 1993; Galama et al. 1998; Stanek et al. 2003; Malesani et al. 2004; Modjaz et al. 2006; Pian et al. 2006), a little star formation rate in a nearby host galaxy, possibly associated with kilonova, as well as the GW radiation of some short-duration GRBs (Berger et al. 2013; Tanvir et al. 2013; Jin et al. 2016; Abbott et al. 2017; Lü et al. 2017; Lamb et al. 2019; Troja et al. 2019; Troja 2023), suggest that the majority of long- and short-duration GRBs are likely related to the deaths of massive stars (type II) and mergers of two compact stellar objects (type I), respectively (Woosley 1993; Paczyński 1998; Zhang 2006; Zhang et al. 2009; Lü et al. 2010; Metzger 2019).

However, the measurement of $T_{90}$ is energy and instrument dependent (Qin et al. 2013), and a long-duration GRB may be disguised as a short-duration GRB due to the tip-of-iceberg effect (Lü et al. 2014). From an observational point of view, several peculiar GRBs break the traditional understanding of GRB progenitors, namely, that long- and short-duration GRBs are originated from mergers of two compact stars and death of massive stars, respectively, e.g., GRB 200826A (Ahumada et al. 2021; Zhang et al. 2021; Rossi et al. 2022), GRB 060614 (Gehrels et al. 2006; Yang et al. 2015), GRB 210704A (Becerra et al. 2023), GRB 211211A (Rastinejad et al. 2022; Troja et al. 2022; Yang et al. 2022; Chang et al. 2023; Gompertz et al. 2023), and GRB 211227A (Lü et al. 2022; Ferro et al. 2023). Lü et al. (2010) proposed a new phenomenological classification method of GRBs by introducing a new parameter $\varepsilon = E_{\gamma,{\rm iso},52}/E_{p,z,2}^{5/3}$, where $E_{\gamma,{\rm iso}}$ and $E_{p,z}$ are the isotropic gamma-ray energy and the cosmic rest-frame spectral peak energy, respectively. On the other hand, probably the most definite criterion to differentiate type I GRBs from type II GRBs is through detecting their GW signals by advanced LIGO/Virgo, and it opened a new window into the study of the properties of GRB progenitors (Berger 2014), such as the first direct detection of a GW event (GW170817) associated with short GRB 170817A from neutron star–neutron star (NS–NS) merger (Abbott et al. 2017; Goldstein et al. 2017; Savchenko et al. 2017; Zhang et al. 2018).

Whatever the collapsar of massive stars or mergers of two compact stars, a hyper-accretion black hole or a rapidly spinning, strongly magnetized neutron star (millisecond magnetar) as the central engine may be formed, and it (the central engine) launches a relativistic outflow (Usov 1992; Dai & Lu 1998; Popham et al. 1999; Narayan et al. 2001; Zhang & Mészáros 2001; Metzger et al. 2010; Lei et al. 2013; Lü & Zhang 2014; Lü et al. 2015). However, how to identify the central engines of GRBs remains an open question (Zhang 2011). The evidence of observed plateau emission component or an extremely steep drop following the plateau (known as internal plateaus) in the X-ray afterglows is suggested to be originated from a millisecond magnetar central engine (Troja et al. 2007; Lyons et al. 2010; Rowlinson et al. 2010, 2013; Lü & Zhang 2014; Lü et al. 2015). On the other hand, the released energy of GRB exceeded the energy budget of magnetar, or the later giant bump in the afterglow, supporting evidence of the black hole as the central engine of GRB (Kumar et al. 2008; Wu et al. 2013; Zhao et al. 2021).

Most recently, an extremely bright and peculiar long-duration GRB 230307A that triggered the Fermi Gamma-ray Burst Monitor (GBM; Dalessi & Fermi GBM Team 2023), Gravitational Wave High-energy Electromagnetic Counterpart All-sky Monitor (GECAM; Xiong et al. 2023), and Konus-Wind (Svinkin et al. 2023) attracted attention with redshift $z = 0.065$ (Dichiara et al. 2023; Levan et al. 2023; Sun et al. 2023; Yang et al. 2023). The burst was also observed by several other missions and optical telescopes including AGILE (Casentini et al. 2023), James Webb Space Telescope (JWST; Levan et al. 2023), and Lobster Eye Imager for Astronomy (LEIA) in the soft X-ray band (Sun et al. 2023). The burst's duration ($T_{90}$) is about 42 s in the 10–1000 keV energy range, and the peak flux and total fluence in the energy range are $4.48 \times 10^{-4}$ erg cm$^{-2}$ s$^{-1}$ and $3 \times 10^{-3}$ erg cm$^{-2}$, respectively (Sun et al. 2023). More interestingly, no associated supernova signature is detected for GRB 230307A, even down to very stringent limits at such a low redshift, but a possible associated kilonova is observed by several optical telescopes (Dichiara et al. 2023; Levan et al. 2023; Yang et al. 2023). That observed evidence together with heavy element nucleosynthesis suggests that GRB 230307A is originated from a binary compact star merger (Dichiara et al. 2023; Du et al. 2024; Levan et al. 2023; Yang et al. 2023), but this remains in debate. In particular, Sun et al. (2023) discovered a soft X-ray emission (observed by LEIA 0.5 4 keV) that exhibits a plateau emission ($\sim t^{-0.4}$) smoothly connected with a $\sim t^{-2.33}$ segment at a broken time $t_b \simeq 80$ s, and its spectrum does not show spectral evolution during the first 100 s. Sun et al. (2023) proposed that the soft X-ray emission observed by LEIA is powered by the magnetar central engine with magnetic dipole radiation. However, the efficiency from the spin-down luminosity of magnetar to X-ray emission is strongly dependent on the injected luminosity of the magnetar (Xiao & Dai 2017), and it is quite simple and inaccurate to adopt the magnetar model to directly fit the observed X-ray data.

In this paper, we attempt to identify the progenitor of GRB 230307A by comparing with other type I and type II GRBs and test whether it is a possible tip-of-iceberg effect (in Section 2). The main novel point of this paper is that we consider the varying efficiency from the spin-down luminosity of the magnetar to X-ray emission, which is related to the radiations. It is more physical and more accurate to estimate parameters of the magnetar by comparing constant efficiency. Then, we derive the parameters of the magnetar, which is the central engine of GRB 230307A. The method is quite different from previous works, which adopted a constant efficiency (in Section 3). Conclusions are drawn in Section 4 with some additional discussions. Throughout the paper, we use the notation $Q = 10^n Q_n$ in CGS units and adopt a concordance cosmology with parameters $H_0 = 71$ km s$^{-1}$ Mpc$^{-1}$, $\Omega_M = 0.30$, and $\Omega_\Lambda = 0.70$.

## 2. Progenitor and Classification

With the measured $z = 0.065$ (corresponding to luminosity distance $D_L \sim 292$ Mpc), GRB 230307A has a rest-frame duration $T_{90}/(1 + z)$, much longer than 2 s. Generally speaking, it should be in the category of long-duration type II GRBs, which are from the collapse of a massive star (Zhang et al. 2009). However, no associated supernova signature is detected for GRB 230307A, even down to very stringent limits at such a low redshift, but a possible signature of an associated kilonova is observed by several optical telescopes (Dichiara et al. 2023; Levan et al. 2023; Yang et al. 2023). This naturally raises the interesting question regarding the progenitor system of this burst.

Lü et al. (2010) proposed a new phenomenological classification method of GRBs by introducing a new parameter $\varepsilon = E_{\gamma,\mathrm{iso},52}/E_{p,z,2}^{5/3}$, and a clear bimodal distribution of $\varepsilon$ with a separation $\varepsilon = 0.03$ is found to correspond to high $-\varepsilon$ and low $-\varepsilon$, respectively. Here, $E_{\gamma,\mathrm{iso},52}$ is the isotropic burst energy in units of $10^{52}$ erg, and $E_{p,z,2}$ is the rest-frame peak energy of the $\nu f_\nu$ spectrum of the prompt gamma-rays in units of $10^2$ keV. They found that the high $-\varepsilon$ region encloses the typical long-duration GRBs with high luminosity, high-z "rest-frame-short" GRBs (such as GRBs 090423 and 080913), as well as high-z short GRBs (such as GRB 090426), which possibly correspond to the death of massive stars (type II origin), while all the GRBs that are argued to be of (type I origin) compact star mergers are found to be clustered in the low $-\varepsilon$ region. Following this method, we also calculate the $\varepsilon$ of GRB 230307A by adopting the values of $E_{\gamma,\mathrm{iso}}$ and $E_{p,z}$ measured by GECAM (Sun et al. 2023). We find that the value of $\varepsilon$ for this burst is about 0.05, which is larger than that of all type I GRBs and close to the separated line of $\varepsilon$ distribution (see Figure 1), but out of the $3\sigma$ of the Type II GRBs distribution. It suggests that the GRB 230307A seems to be with an ambiguous progenitor, either one type of compact star merger (e.g., neutron star–neutron star (NS–NS), neutron star–white dwarf (NS–WD), or neutron star–black hole (NS–BH)), which is different from other typical short GRBs, or this method at least is not suitable for applying to GRB 230307A to do the classification. Moreover, GRBs 211211A, 211227A, and 210704A are believed to be from compact star mergers (Lü et al. 2022; Rastinejad et al. 2022; Troja et al. 2022; Yang et al. 2022; Becerra et al. 2023; Chang et al. 2023; Ferro et al. 2023). In order to compare with the two peculiar long-duration GRBs, we also calculate the values of $\varepsilon$ for those two GRBs and find that the $\varepsilon$ values of those two GRBs are quite similar to that of GRB 230307A. It implies that the compact star mergers may produce both long-duration (such as GRBs 230307A, 211211A, and GRB 211227A) and short-duration GRBs, but the physical origins remain unknown.

Next, Lü et al. (2014) proposed to add a third dimension, "amplitude," into consideration to classify GRBs by using the prompt gamma-ray data. The motivation is to study the possibility that a real long GRB may be observed as a "short" one if the majority of prompt emission of the GRB is too faint to be detected above the background. They called this the effect of "tip of iceberg." The definition of an effective amplitude parameter is $f_{\mathrm{eff}} = F_p'/F_B$, where $F_p'$ is the simulated peak flux of prompt emission, which is rescaled down for multiplying by a factor from an original long-duration GRB light curve until its signal above the background has a duration $T_{90,\mathrm{eff}}$ just shorter than 2 s. $F_B$ is the flux of the background. They found that most short GRBs are likely not the "tip of iceberg" of long GRBs. Note that the value of $f_{\mathrm{eff}}$ can only tell us whether one short-duration GRB is intrinsically short, likely produced by a massive star, and it cannot directly tell us the progenitor. Thus, we perform the same analysis for GRB 230307A following Lü et al. (2014) and measured both the $F_p'$ and $F_B$ of GRB 230307A. We find that the value of $f_{\mathrm{eff}}$ is 1.23, which is less than most short-duration GRBs likely produced by a compact star merger, but it (red star) is comparable with that of typical type II GRBs (gray) as shown in Figure 1. Moreover, we also calculate the probability ($p$) of this being a disguised type I GRB as $p > 0.22$, and it suggests that GRB 230307A is not likely to be from the "tip of iceberg" effect of a long-duration

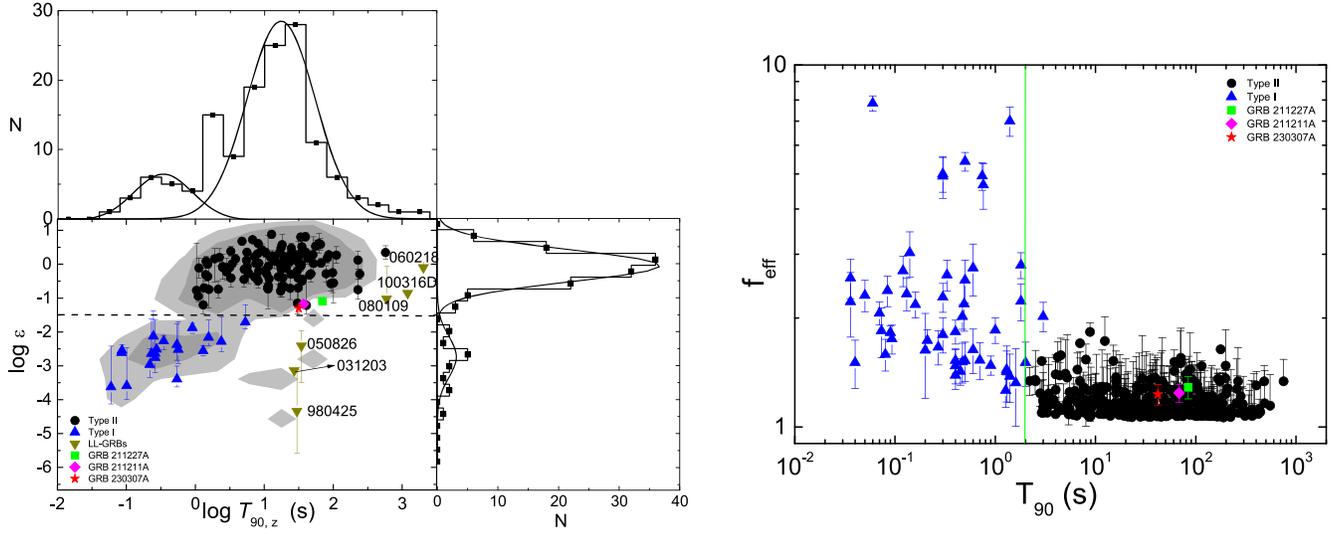

**Figure 1.** Left: 1D and 2D distributions of GRBs samples in $T_{90}-\varepsilon$ space. The dashed line is $\varepsilon = 0.03$. Right: $f_{\rm eff} - T_{90}$ for type I and type II GRBs, which are from Lü et al. (2014). The vertical solid line is $T_{90} = 2$ s. Blue triangles and black solid circles represent the type I and type II GRB candidates, and dark yellow triangles denote the nearby low-luminosity long GRBs. GRBs 230307A, 211211A, and 211227A correspond to red star, pink diamond, and green square, respectively.

GRB. We also calculate the values of $f_{\rm eff}$ and the probability for GRBs 211211A and 211227A for comparison and find that the values of both $f_{\rm eff}$ and probability of GRB 230307A are consistent with that of GRBs 211211A and 211227A.

Sun et al. (2023) compared some empirical relationships of GRB 230307A with other type I and type II GBRs, such as MVT−$T_{90}$, $E_{\rm p} - E_{\rm iso}$, and $L_{\gamma,\rm iso} - \tau_{\rm s}$. Here, MVT and $\tau_{\rm s}$ are the minimum variability timescale and spectral lag. They find that GRB 230307A is more likely to be consistent with that of type I GRB despite being a long-duration GRB. The results of $\epsilon$ classification and the possible signature of an associated kilonova suggest that GRB 230307A is more likely to be from the merger of compact stars, but it is different from other typical type I GRBs. We suspect that GRB 230307A, together with GRBs 211211A and 211227A, are from one type of two compact star mergers (e.g., NS–NS, NS–WD, or NS–BH).

### 3. Magnetar Central Engine

Magnetars as the central engine of both long- and short-duration GRBs is extensively discussed (Dai & Lu 1998; Zhang & Mészáros 2001; Fan & Xu 2006; Troja et al. 2007; Rowlinson et al. 2010; Dall'Osso et al. 2011; Metzger et al. 2011; Rowlinson et al. 2013; Lü & Zhang 2014; Lü et al. 2015; Sarin et al. 2019). A soft X-ray emission (observed by LEIA 0.54 keV) of GRB 230307A was discovered to exhibit a plateau emission ($\sim t^{-0.4}$) smoothly connected with a $\sim t^{-2.33}$ segment at a broken time $t_b \simeq 80$ s, and it suggests that a millisecond magnetar maybe reside in the central engine to power the plateau emission (Sun et al. 2023). In this section, we consider varying efficiency (from the spin-down luminosity of magnetar to X-ray emission), which is strongly dependent on the injected luminosity of the magnetar, and find out whether the parameters of the magnetar for GRB 230307A fall into a reasonable range.

#### 3.1. Magnetar Central Engine and Synchrotron Radiation

The total rotation energy of a millisecond magnetar can be expressed as

$$E_{\rm rot} = \frac{1}{2} I \Omega^2 \simeq 2 \times 10^{52} \text{ erg } M_{1.4} R_6^2 P_{-3}^{-2}, \quad (1)$$

where $I$ is the moment of inertia; and $\Omega$, $P$, $R$, and $M$ are the angular frequency, rotating period, radius, and mass of the neutron star, respectively. In general, the magnetar loses its rotational energy via electromagnetic ($L_{\rm EM}$) and gravitational-wave ($L_{\rm GW}$) radiations (Zhang & Mészáros 2001; Fan et al. 2013; Giacomazzo & Perna 2013; Lasky & Glampedakis 2016):

$$-\frac{dE_{\rm rot}}{dt} = -I\Omega\dot{\Omega} = L_{\rm total} = L_{\rm EM} + L_{\rm GW}$$
$$= \frac{B_{\rm p}^2 R^6 \Omega^4}{6c^3} + \frac{32 G I^2 \epsilon^2 \Omega^6}{5c^5}, \quad (2)$$

where $B_{\rm p}$ is the surface magnetic field at the pole and $\epsilon = 2(I_{xx}I_{yy})/(I_{xx}+I_{yy})$ is the ellipticity describing how large the neutron star deformation is. Its angular frequency ($\Omega$) evolution can be described with the torque equation as the following:

$$\dot{\Omega} = d\Omega/dt = -k\Omega^n, \quad (3)$$

where $\Omega = \Omega(t) = 2\pi/P(t)$. $k$ and $n$ are a constant of proportionality and braking index of magnetar, respectively. The solution of Equation (1) can be expressed as (Zhang & Mészáros 2001; Lasky et al. 2017; Lü et al. 2019)

$$\Omega(t) = \Omega_0 \left(1 + \frac{t}{\tau}\right)^{\frac{1}{1-n}}, \quad (4)$$

where $\Omega_0$ is the initial angular frequency at $t=0$ and $\tau = \Omega_0^{1-n}/((n-1)k)$ is the spin-down timescale. The luminosity of the magnetic dipole torque is determined by $L_{\rm EM} = B_{\rm p}^2 R^6 \Omega^4/6c^3$. Therefore, the observed X-ray luminosity can be written as

$$L_{\rm X,obs} = (1+z)\eta L_{\rm EM} = (1+z)\eta L_0 \left(1 + \frac{t}{\tau}\right)^{\frac{4}{1-n}}, \quad (5)$$

where $L_0 = B_{\rm p}^2 R^6 \Omega_0^4/6c^3 \approx 10^{49}$ erg s$^{-1} B_{15}^2 R_6^6 P_{0,-3}^{-4}$ is the spin-down luminosity of the magnetar. $\eta$ is the conversion efficiency from the magnetar luminosity to the X-ray emission, and it varies with the injected luminosity of the magnetar (Xiao et al. 2019). For a given X-ray energy range, $\eta$ is defined as

$$\eta = \frac{\int_{0.5\,\text{keV}}^{4\,\text{keV}} L_\nu d\nu}{L_{\text{EM}}}. \quad (6)$$

In practice, the energy of the dipole radiation can propagate outward with a Poynting-flux–dominated outflow. We assume that the energy spectrum of accelerated electrons is roughly a power-law distribution (Sironi & Spitkovsky 2014; Guo et al. 2015; Werner et al. 2016),

$$N(\gamma_e) d\gamma_e \propto \gamma_e^{-p} d\gamma_e, \ \gamma_e \geqslant \gamma_m, \quad (7)$$

where $\gamma_e$ and $\gamma_m$ are the Lorentz factor and minimum Lorentz factor of accelerated electrons, respectively (Beniamini & Giannios 2017). $p$ is the power-law index of accelerated electrons, and $p = 4\sigma^{-0.3}$ is adopted in our calculations according to a reasonable fit for the results of numerical calculations. Here, $\sigma$ is the magnetization parameter of the outflow. The Poynting luminosity at a radius ($r_p$) is written as

$$L_p = \frac{c(r_p B_{\text{jet}})^2}{4\pi} = L_{\text{EM}} \left(1 - \frac{\Gamma}{\Gamma_{\text{sat}}}\right), \quad (8)$$

where $B_{\text{jet}}$ and $\Gamma$ are the jet magnetic field strength and bulk Lorentz factor within the outflow, respectively. $\Gamma_{\text{sat}}$ is the bulk Lorentz factor at saturation radius ($r_{\text{sat}}$) of the outflow.

By considering the synchrotron radiation, one defines a parameter $\gamma_c$, which is the critical Lorentz factor for electron synchrotron radiation, so that the radiated spectrum can be separated as two different regions by comparing with $\gamma_c$, e.g., fast-cooling ($\gamma_m > \gamma_c$) and slow-cooling ($\gamma_m < \gamma_c$) cases (Sari et al. 1998). We list the luminosity ($L_\nu$) at each frequency ($\nu$) as follows for these two cases:

1. Fast-cooling regime:

$$L_\nu = \begin{cases} L_{\nu,\text{max}} \left(\frac{\nu}{\nu_c}\right)^{\frac{1}{3}}, & \nu < \nu_c \\ L_{\nu,\text{max}} \left(\frac{\nu}{\nu_c}\right)^{-\frac{1}{2}}, & \nu_c < \nu < \nu_m \\ L_{\nu,\text{max}} \left(\frac{\nu_m}{\nu_c}\right)^{-\frac{1}{2}} \left(\frac{\nu}{\nu_m}\right)^{-\frac{p}{2}}, & \nu_m < \nu < \nu_{\text{max}} \end{cases} \quad (9)$$

2. Slow-cooling regime:

$$L_\nu = \begin{cases} L_{\nu,\text{max}} \left(\frac{\nu}{\nu_m}\right)^{\frac{1}{3}}, & \nu < \nu_m \\ L_{\nu,\text{max}} \left(\frac{\nu}{\nu_m}\right)^{-\frac{p-1}{2}}, & \nu_m < \nu < \nu_c \\ L_{\nu,\text{max}} \left(\frac{\nu_c}{\nu_m}\right)^{-\frac{p-1}{2}} \left(\frac{\nu}{\nu_c}\right)^{-\frac{p}{2}}, & \nu_c < \nu < \nu_{\text{max}} \end{cases} \quad (10)$$

where $L_{\nu,\text{max}}$, $\nu_m$, $\nu_c$, and $\nu_{\text{max}}$ correspond to the maximal luminosity, typical frequency, cooling frequency, and the maximal frequency, respectively. The formulas of $\nu_m$, $\nu_c$, and $L_{\nu,\text{max}}$ are expressed as (Sari et al. 1998)

$$\nu_c = \frac{1}{1+z} \frac{72\pi e m_e c^3 \Gamma^6}{\sigma_T^2 B_{\text{jet}}^3 r^2}, \quad (11)$$

$$\nu_m = \frac{1}{1+z} \gamma_m^2 \frac{eB_{\text{jet}}}{2\pi m_e c}, \quad (12)$$

$$L_{\nu,\text{max}} = (1+z) \frac{m_e c^2 \sigma_T B_{\text{jet}} N_e}{3e}$$
$$= (1+z) \frac{m_e c^2 \sigma_T N_e}{3 e r_p} \left[\frac{4\pi L_{\text{EM}}}{c}\left(1 - \frac{\Gamma}{\Gamma_{\text{sat}}}\right)\right]^{1/2}, \quad (13)$$

where $\sigma_T$, $e$, and $N_e$ are the Thomson scattering cross section, the electron charge, and the total number of emitting electrons in the jet at $r_p$, respectively.

### 3.2. Magnetar Model Fits and Parameter Constraints

Based on Equations (9) and (10), the spectrum of synchrotron radiation depends on the radius, and we integrate it from photospheric radius to the saturation radius. Together with Equation (5), one has to obtain the value of $\Gamma_{\text{sat}}$, which is related to $\sigma_0$, namely, $\Gamma_{\text{sat}} = \sigma_0^{3/2}$. However, the conditions for determining the initial magnetization parameter $\sigma_0$ is unclear and challenging to detect. Therefore, we adopt some typical values as suggested by Xiao & Dai (2017), e.g., $\Gamma_{\text{sat}} = 10^2$, $10^{2.5}$, $10^3$, $10^4$, and $10^5$. More details can be found in Beniamini & Giannios (2017) and Xiao & Dai (2017). On the other hand, the X-ray radiation efficiency ($\eta$) is strongly dependent on the injected luminosity ($L_{\text{EM}}$) of the magnetar (Xiao & Dai 2017). For simplification, we adopt $L_{\text{EM}} \approx L_0$ to do the calculations.

The soft X-ray data from 0.5 to 4 keV are taken from Sun et al. (2023). Then, we take ($L_0$, $n$, $\tau$) as parameters and adopt the Python package EMCEE (Foreman-Mackey et al. 2013) to fit the data. Figure 2 shows an example of the soft X-ray light-curve fits by the magnetar model for $\Gamma_{\text{sat}} = 10^4$ and 2D histograms with parameter constraints. The best-fitting parameters are shown in Table 1 for given different $\Gamma_{\text{sat}}$.

By adopting the best-fitting parameters above, we can infer the basic parameters of the magnetar, e.g., initial magnetic field $B_p$ and period $P_0$. Since a jet break was not detected in this GRB, we have to adopt the jet opening angle $\theta_j = 15°$ with beaming factor $f_b = 0.034$ to do the correction (Rowlinson et al. 2013). From a theoretical point of view, the braking index ($n$) of magnetar is closed to 3 when its rotation energy loss is dominated by electromagnetic radiation. However, in our fitting, $n$ is much less than 3, suggesting that at least another braking mechanism exists, such as fallback accretion onto the magnetar (Metzger et al. 2018) and other braking mechanisms we do not know. In any case, in our estimation, we do not consider other braking mechanisms. The inferred timescale is no longer than the spin-down timescale, which was solely determined by electromagnetic torque $\tau_{\text{EM}}$, one of which has $\tau \lesssim \tau_{\text{EM}} = 2 \times 10^3 \, s \, I_{45} B_{15}^{-2} R_6^{-6} P_{0,-3}^2$. We employ typical values of $R_6 \sim 1$ km and $I_{45} \sim 1.9$ g cm$^2$ of the magnetar to calculate the maximum values of the spin period ($P$) and magnetic field strength ($B_p$; Piro et al. 2017), and the estimated upper limits of $P$ and $B_p$ are presented in Table 1 for given different $\Gamma_{\text{sat}}$.

One interesting question is whether the observations of GRB 230307A obey synchrotron radiation. In Figure 3, we also plot the spectrum of synchrotron radiation to compare with the observations.[3] It is found that the model prediction can pass through the observational area. It implies that synchrotron radiation is consistent with the observations of GRB 230307A. Note that $\Gamma_{sat} = 10^2$ corresponds to $r_{ph} > r_{sat}$, meaning that the radiation is dominated by photosphere emission without synchrotron emission. Therefore, we neglect the case of $\Gamma_{sat} = 10^2$ in Figure 3.

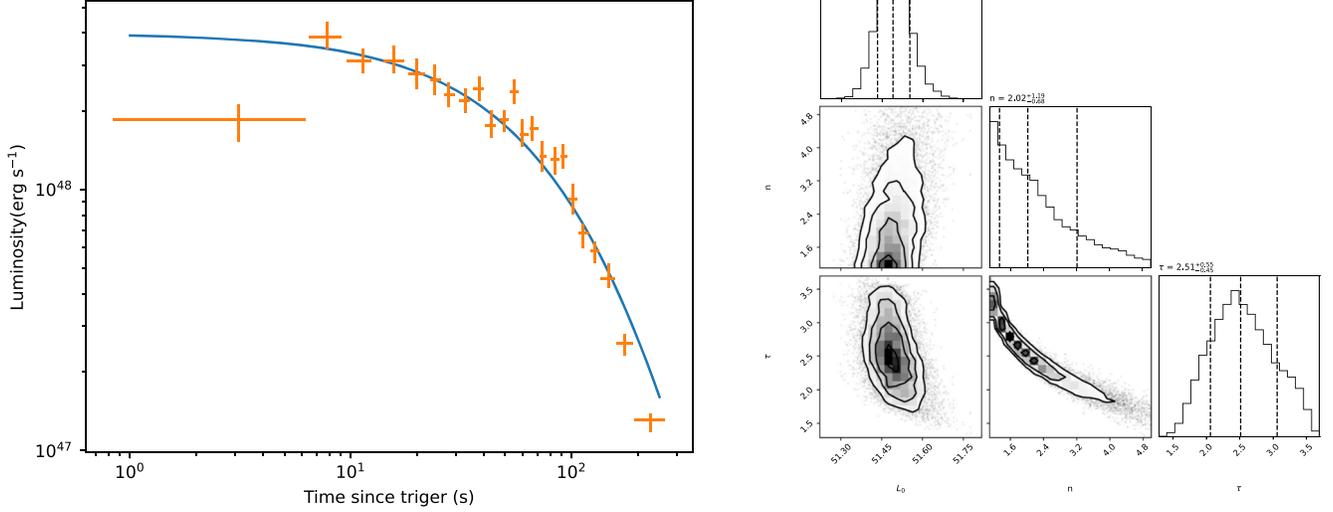

**Figure 2.** Left: the soft X-ray light curve of GRB 230307A (orange points), as well as the fits by magnetar model for $\Gamma_{sat} = 10^4$ (blue line). Right: 2D histograms and parameter constraints of magnetar model fit for soft X-ray light curve of GRB 230307A with $\Gamma_{sat} = 10^4$ case.

**Table 1**
The Best-fitting Results and Estimated Upper Limits of $P$ and $B_p$ of a Magnetar for Given Different $\Gamma_{sat} = 10^2, 10^{2.5}, 10^3, 10^4,$ and $10^5$

| $\Gamma_{sat}$ | $\log L_0$ (erg s$^{-1}$) | $n$ | $\log \tau$ (s) | $P$ (ms) | $B_{p,15}$ (G) |
|---|---|---|---|---|---|
| $\Gamma_{sat} = 10^2$ | $53.62^{+0.23}_{-0.2}$ | $1.47^{+0.47}_{-0.26}$ | $2.32^{+0.44}_{-0.46}$ | $0.11^{+0.13}_{-0.06}$ | $0.48^{+1.27}_{-0.35}$ |
| $\Gamma_{sat} = 10^{2.5}$ | $52^{+0.22}_{-0.18}$ | $1.57^{+0.58}_{-0.33}$ | $2.3^{+0.49}_{-0.47}$ | $0.74^{+0.83}_{-0.42}$ | $3.25^{+8.56}_{-2.44}$ |
| $\Gamma_{sat} = 10^3$ | $50.8^{+0.1}_{-0.1}$ | $1.75^{+0.86}_{-0.47}$ | $2.44^{+0.5}_{-0.47}$ | $2.52^{+2.34}_{-1.26}$ | $9.38^{+21.68}_{-6.74}$ |
| $\Gamma_{sat} = 10^4$ | $51.49^{+0.06}_{-0.06}$ | $2.02^{+1.19}_{-0.68}$ | $2.51^{+0.55}_{-0.45}$ | $1.05^{+0.84}_{-0.53}$ | $3.61^{+7.29}_{-2.66}$ |
| $\Gamma_{sat} = 10^5$ | $54.13^{+0.09}_{-0.09}$ | $1.74^{+0.91}_{-0.47}$ | $2.49^{+0.5}_{-0.47}$ | $0.05^{+0.05}_{-0.03}$ | $0.18^{+0.41}_{-0.13}$ |

For the derived of $P$ and $B_p$ for given $\Gamma_{sat}$, we find that spin periods of the magnetar for $\Gamma_{sat} = 10^2, 10^{2.5},$ and $10^5$ are so rapid that they exceed the spin period by less than 1 ms (through numerical simulations) for a neutron star (Lattimer & Prakash 2004). Taking into account the reasons discussed above, we restrict our consideration to $\Gamma_{sat} \sim 10^3$–$10^4$ as the most plausible range for understanding the magetar wind. The range of $\Gamma_{sat}$ in this work is consistent with that of Xiao et al. (2019).

In order to compare the estimated $P$ and $B_p$ of GRB 230307A in this work with that of other type I GRBs derived by internal plateau in X-ray afterglow (Lü et al. 2015), Figure 4 shows the $B_p$–$P$ diagram for GRB 230307A with $\Gamma_{sat} = 10^3$ and $10^4$. It seems that the upper limit of estimated $P$ and $B_p$ of GRB 230307A is lower than other typical type I GRB samples but fall into a reasonable range.

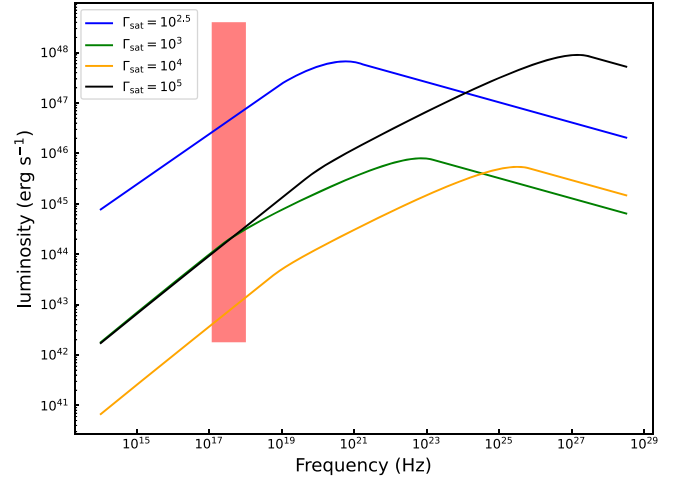

**Figure 3.** Synchrotron radiation spectrum with different $\Gamma_{sat}$. The observed luminosity between 0.5 and 4 keV is shown in the shadowed area.

## 4. Conclusion and Discussion

GRB 230307A is an extremely bright and peculiar long-duration GRB with a duration of ∼42 s detected by GECAM (Xiong et al. 2023) and Fermi (Dalessi & Fermi GBM Team 2023), as well as Konus-Wind (Svinkin et al. 2023), with a measured redshift of $z = 0.076$ (Dichiara et al. 2023; Levan et al. 2023; Sun et al. 2023; Yang et al. 2023). No associated supernova signature is detected for GRB 230307A, not even at very stringent limits at a low redshift, but a possible associated kilonova is observed by several optical telescopes (Dichiara et al. 2023; Levan et al. 2023; Sun et al. 2023; Yang et al. 2023). The observed evidence together with heavy element nucleosynthesis suggests that GRB 230307A may be originated from a binary compact star merger (Dichiara et al. 2023;

---

[3] The peak flux of upper and lower limits in the energy range of (0.5–4) keV are ∼3.65 × 10$^{-7}$ erg cm$^{-2}$ s$^{-1}$ and ∼1.78 × 10$^{-13}$ erg cm$^{-2}$ s$^{-1}$, respectively (Sun et al. 2023).

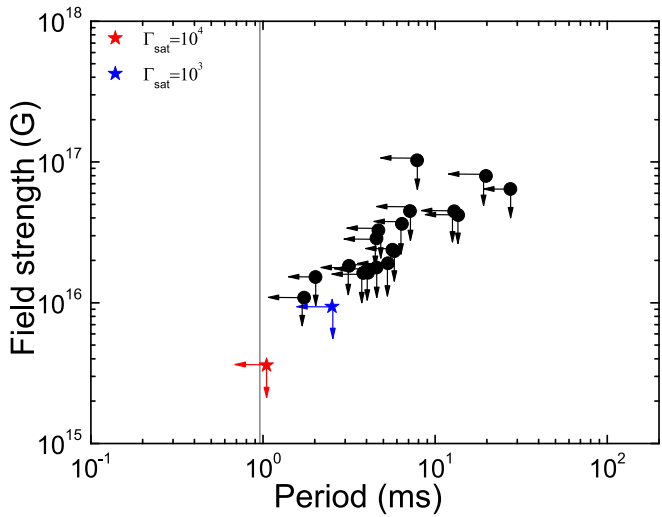

**Figure 4.** Inferred magnetar parameters and initial spin period $P$ vs. surface polar cap magnetic field strength $B_p$ derived for short GRBs with internal plateau (black dots) and GRB 230307A (star) with $\Gamma_{sat} = 10^3$ and $10^4$. The vertical solid line is the breakup spin period limit for a neutron star (Lattimer & Prakash 2004).

Du et al. 2023; Levan et al. 2023; Sun et al. 2023; Yang et al. 2023) but this remains in debate.

A soft X-ray emission (observed by LEIA 0.54 keV) of GRB 230307A was discovered to exhibit a plateau emission ($\sim t^{-0.4}$) smoothly connected with a $\sim t^{-2.33}$ segment at a broken time $t_b \simeq 80$ s, and it suggests that a millisecond magnetar may reside in the central engine to power the plateau emission (Sun et al. 2023). In this paper, we attempt to identify the progenitor of GRB 230307A and investigate the magnetar central engine model with varying efficiency. We find the following interesting results.

1. The calculated values of $\varepsilon \sim 0.05$ and $f_{eff} = 1.23$ for GRB 230307A, together with empirical relationships and the possible signature of an associated kilonova, suggest that GRB 230307A is not likely to be from the "tip of iceberg" of a long-duration GRB but is more likely to be from mergers of compact stars, e.g., merger of NS–NS or NS–WD.

2. The soft X-ray emission of GRB 230307A is consistent with energy injection from the magnetar central engine for varying efficiency. The derived magnetar surface magnetic field $B_p$ and the initial spin period $P$ fall into a reasonable range, i.e., $B < 9.4 \times 10^{15}$ G, $P < 2.5$ ms for $\Gamma_{sat} = 10^3$ and $B < 3.6 \times 10^{15}$ G, $P < 1.05$ ms for $\Gamma_{sat} = 10^4$, respectively.

The observational evidence of some long-duration GRBs, at least for GRBs 060614, 211211A, 211227A, and 230307A, as well as some typical short-duration GRBs, point to the fact that compact binary mergers can power both long- and short-duration GRBs. Recently, Gottlieb et al. (2023) proposed a unified picture of compact binary mergers to power both long- and short-duration GRBs via numerical simulations. They found that the duration of GRBs is dependent on the total mass of the binary, the mass of the disk, and the mass ratio of the binary. However, observational evidence still need to be found to support this hypothesis.

The first direct detection of a GW event (GW170817) associated with short GRB 170817A opened a new window to study the properties of a progenitor for type I GRBs (Abbott et al. 2017; Goldstein et al. 2017; Savchenko et al. 2017; Zhang et al. 2018). Yin et al. (2023) presented more detailed calculations for the GW signals from mergers of NS–BH and NS–WD. The GW signals from the mergers of NS–NS is quite different from those of the mergers of NS–WD, and it suggests that the GW signals from the merger of compacts can be used as a probe to distinguish the progenitor of GRB 230307A-like events in the feature. Multi-messenger observations of similar events (e.g., GRB 211211A-like and GRB 211227A-like events) hold the promise of eventually unveiling the identity of the progenitor of these peculiar systems.

## Acknowledgments

We thank Haoyu Yuan and Di Xiao for helpful discussions of the radiations. This work is supported by the Guangxi Science Foundation the National (grant Nos. 2023GXNSFDA026007, and 2017GXNSFFA198008), the Natural Science Foundation of China (grant Nos. 11922301 and 12133003), and the Program of Bagui Scholars Program (LHJ).

## ORCID iDs

ZhaoWei Du ⓘ https://orcid.org/0000-0002-4375-3737
HouJun Lü ⓘ https://orcid.org/0000-0001-6396-9386
EnWei Liang ⓘ https://orcid.org/0000-0002-7044-733X